# Title: Hot wires and film boiling: Another look at carbonyl formation in electronic cigarettes

**Authors:** S. Talih[1], R. Salman[1], E. Karam[1], M. El-Hourani[1], R. El-Hage[2], N. Karaoghlanian[1], A. El-Hellani[2], N. Saliba[2], A. Shihadeh[1]*

**Affiliations:**

[1] Mechanical Engineering Department, Faculty of Engineering and Architecture, American University of Beirut, Bliss Street, P.O. Box 11-0236, Beirut, Lebanon.

[2] Chemistry Department, Faculty of Arts and Sciences, American University of Beirut, Bliss Street, P.O. Box 11-0236, Beirut, Lebanon

*Correspondence to Alan Shihadeh, as20@aub.edu.lb

**Abstract:** Electronic cigarettes (ECIGS) are a class of tobacco products that emit a nicotine-containing aerosol by heating and vaporizing a liquid. Apart from initiating nicotine addiction in nonsmokers, a persistent concern about these products is that their emissions often include high levels of carbonyl species, toxicants thought to cause most non-cancer pulmonary disease in smokers. Carbonyls are known thermal degradation products of the primary ECIG liquid constituents: propylene glycol (PG) and vegetable glycerin (VG). To date, there is no method available to predict the operating conditions for which an ECIG of a given design will emit large quantities of carbonyls. This study examined whether the phenomenon of film boiling can account for observations of high carbonyl emissions under certain operating conditions, and if so, whether film boiling theory can be invoked to predict conditions where such emissions are likely. We measured the critical heat flux for several common heating materials and liquids, and carbonyl emissions for several ECIG types while varying power. We found that emissions rise drastically whenever power exceeds the value corresponding to the critical heat flux. While limiting the heat flux to below this threshold can greatly reduce carbonyl exposure, ECIG manufacturer operating instructions often exceed it. Product regulations that limit heat flux may reduce the public health burden of electronic cigarette use.

**One Sentence Summary:** Carbonyl emissions from electronic cigarettes are predictable and can be reduced via regulatory limits on energy flux.

Electronic cigarettes (ECIGs) are a rapidly growing class of tobacco products that contain an electrical heating filament or "coil" and a liquid-saturated fibrous wick (Fig. 1). When the coil is powered, it heats and vaporizes the liquid to produce an inhalable nicotine-containing aerosol mist. The potential benefits and risks of ECIGs to public health are a subject of much debate among scientists and policymakers. On the one hand, ECIG toxicant emissions, may be lower than those of combustible cigarettes (1-5) and may, therefore, reduce disease risk in smokers who switch. On the other hand, ECIGs may initiate nicotine-naïve individuals to a lifetime of addiction and may ultimately induce population-wide increases in combustible tobacco use (6-8). Apart from nicotine addiction, a persistent health concern surrounding ECIG use is that it may expose users to harmful carbonyl compounds (CCs), a powerful class of respiratory toxicants thought to induce the majority of non-cancer pulmonary disease in cigarette smokers (9, 10). Carbonyls, including formaldehyde and acrolein, are produced by thermal degradation of the major ECIG liquid constituents, namely propylene glycol (PG) and vegetable glycerin (VG) (11-14). A key question for product regulation is the degree to which carbonyls and other toxic thermal degradation products can be minimized by product design and operation constraints.

Studies have reported wide ranges of CC emissions, ranging from negligible quantities to several combustible cigarette equivalents in a few puffs (1, 2, 13, 15-20). We have previously shown that electrical power input normalized by heating coil surface area (heat flux, q", $kW/m^2$) predicts CC emissions across devices (16). However, there are ECIG operating conditions in which small increases in power result in disproportionately larger CC emissions (17, 19). We refer to this condition as the "high carbonyl regime" or HCR, during which CC emissions may increase by orders of magnitude. To date, laboratory observations of high CC emissions have been attributed to "dry-puffing", when the ECIG wick runs dry, allowing the coil to reach sufficiently high temperatures that the remaining traces of liquid thermally degrade to form carbonyls (21-23).

"Dry-puffs" may occur when liquid is vaporized by the coil more rapidly than fresh liquid can replenish the wick. However, while studying ECIG toxicant emissions, we have regularly observed HCR onset even when the coil-wick system was well-saturated by liquid. In this study, we examined an alternative construct to understand and predict the onset of HCR, namely a thermo-physical phenomenon known as "film boiling".

Film boiling can occur at the interface of a submerged, heated surface. If the heat flux exceeds a threshold, a thin vapor film forms between the hot surface and the surrounding liquid. When the film forms, it acts as an insulator, impeding heat transfer from the surface, causing surface temperature to rise drastically. The threshold heat flux is called the critical heat flux, or CHF ($kW/m^2$). CHF can be measured or computed theoretically based on liquid properties. If HCR onset is due to film boiling, then CHF can provide a readily computed metric for the probability that a given ECIG design and operating condition will lead to high CC emissions. Potentially, CHF could guide product design regulations for public health ends.

The primary purpose of this study was thus to examine whether CHF predicts the onset of HCR for a given ECIG. To test this theory, we first submerged typical ECIG heating wires that differed by geometry and material and examined whether film boiling occurs in ECIG wire-liquid systems under heat flux conditions that are relevant to ECIG operation. Second, we measured temperature, liquid vaporized, and CC emissions for three different categories of ECIGs over a range of powers and tested whether approaching and exceeding CHF coincides with both HCR onset and behaviors characteristic of film boiling: onset of high-temperature, followed by a limited-vaporization regime in which additional increments in heat flux do not result in greater liquid vaporization rate. Third, we triangulated by varying the ECIG liquid composition and therefore the intrinsic CHF of the wire-liquid combination, and checked whether observed changes in CHF corresponded to predictions from film boiling theory. Finally, using a historical database, we examined whether CHF predicted HCR for a large number of ECIG devices operated under a wide range of conditions.

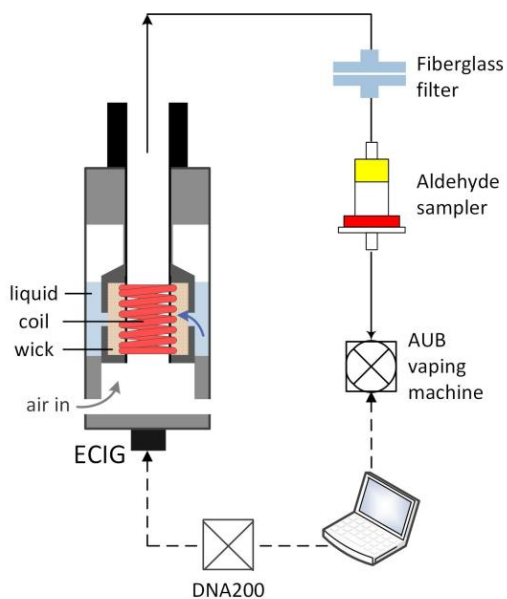

**Fig.1 Schematic of an ECIG and experimental setup**

**Do ECIG wires exhibit film boiling at relevant heat fluxes?**

We investigated whether ECIG coil-liquid systems can transition to film boiling under plausible heat fluxes ($q''$). To do so, we submerged several nickel ECIG wires of varying geometry in PG and powered them in increasing increments while continuously recording their temperatures. nickel wires are commonly used in temperature-controlled ECIG devices because the relatively high temperature coefficient of resistance (TCR) of nickel makes electrical resistance a convenient surrogate measure of mean wire temperature. We observed the temperature record for signs of transition to film boiling and recorded the CHF.

Simultaneously, we visually observed changes to the wire-liquid system as $q''$ increased (Video S1). The visual observations can be described as follows: at low $q''$, we observed convective motion of the liquid and no sign of bubble formation. As we increased $q''$, visible bubbles began to form and detach from the wire surface, indicating a transition from natural convection to nucleate boiling. Nucleate boiling continued and intensified with increasing $q''$ until bubbles began joining to form larger bubbles, and eventually a vapor film over parts of the wire. As we further increased the power, the vapor film grew in length and the wire started to glow red; its recorded temperature increased drastically. This last stage describes the transition to the film boiling regime, where the heating surface is no longer in direct contact with the liquid, and therefore its temperature can depart markedly from the liquid boiling temperature, $T_b$. Previous literature describing boiling processes for submerged wires are similar to this account. (24)

Figure 2B shows an example of the measured temperature (superheat: $\Delta T_{excess} = T - T_b$), for one nickel wire submerged in PG. The data show that with increasing q", a threshold is reached where $\Delta T_{excess}$ jumps, indicating a transition to the film boiling regime (i.e. q" > CHF). Beyond this threshold, $\Delta T_{excess}$ continued to increase monotonically with increasing q". Figure 2C shows that all submerged nickel wires exhibited qualitatively the same behavior. We found that CHF varied between 241 and 474kW/m$^2$, with an mean(SD) of 343(105) kW/m$^2$. A summary of the data is included in Table S3.

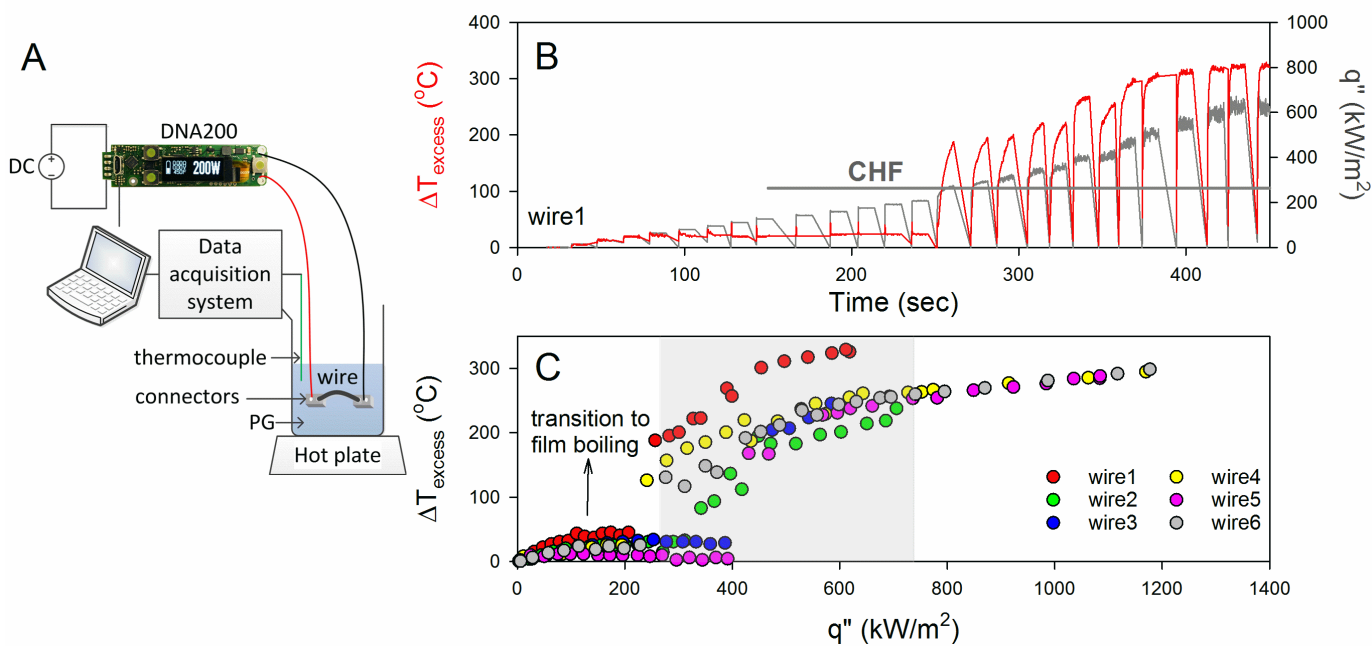

**Fig. 2 Critical heat flux determination for nickel wires submerged in PG**
(A) Schematic of the setup. (B) Example of the superheat profile vs. q" for one wire. (C) Maximum coil superheat as a function of q" for all tested wires; shaded area corresponds to the manufacturer recommended range of operating powers for the VF and SMOK V12-Q4 ECIG devices shown in Figure S1.

We also determined CHF for kanthal and nichrome wires using the same visual observation technique. Interestingly, the data showed variations in CHF across metals, with kanthal exhibiting the highest average CHF of 510(76)kW/m$^2$ followed by nichrome, 390(73)kW/m$^2$, and nickel, 340(105)kW/m$^2$. Apart from nichrome vs. nickel, CHF was significantly different across metals. The difference in CHF across metals has been previously reported and hypothesized to result from differences in TCR; (25) CHF and TCR were reported to be inversely related. Similarly, in this study, the metal with the highest TCR, nickel, (26) resulted in the lowest CHF, and the one with the lowest TCR, kanthal, (27) resulted in the highest CHF.

Importantly, manufacturer-recommended maximum powers for several ECIGs tested here correspond to values exceeding CHF (Table S1). Thus submerged ECIG wires can exhibit film boiling at heat fluxes relevant to ECIG operation.

**Do ECIG devices exhibit evidence of film-boiling at relevant powers?**

We examined temperature response and aerosol emissions while varying power input to three different off-the-shelf models of ECIGs that varied by heating wire material of construction, atomizer geometry, and basic design (see supplementary text S1 for details). One of the three models, the SMOK TF-N2, utilized a nickel heating coil, allowing mean wire temperature to be recorded during puffing.

Figure 3A shows time plots of mean coil temperature measured when individual puffs of 4 sec duration were drawn through the SMOK TF-N2 ECIG device. Plots are arranged in order of increasing power from left to right. The boiling temperature of PG is shown for reference ($T_{sat}$). It can be seen that starting at puff 3 (q"=207 kW/m$^2$), the temperature rapidly reached and then plateaued at $T_{sat}$; subsequent puffs 4-6 showed the same behavior, despite the continuing increase in thermal input. At puff 7 (315 kW/m$^2$), the wire temperature rapidly reached boiling as with the previous puffs, but then continued to rise, albeit at a lower rate. In subsequent puffs at still greater heat inputs, a similar rapid rise to $T_{sat}$ is apparent, however the rate of temperature rise and the final excess temperature continue to increases with increasing heat input (Fig. 3B). The data thus clearly indicate three distinct operating regimes: evaporation (puffs 1-3, q"< 207kW/m$^2$), boiling (puffs 4-6, 207 < q" < 315kW/m$^2$), and superheat (puffs 7 and greater, q"> 315kW/m$^2$). The transition from boiling to superheat regimes appears to occur between 315 kW/m$^2$ and 347 kW/m$^2$; this value is consistent with the mean CHF of 343 kW/m$^2$ that we observed for nickel wires submerged in the same liquid (Fig. 2B and C). This data is consistent with the notion that film boiling can occur in ECIG devices.

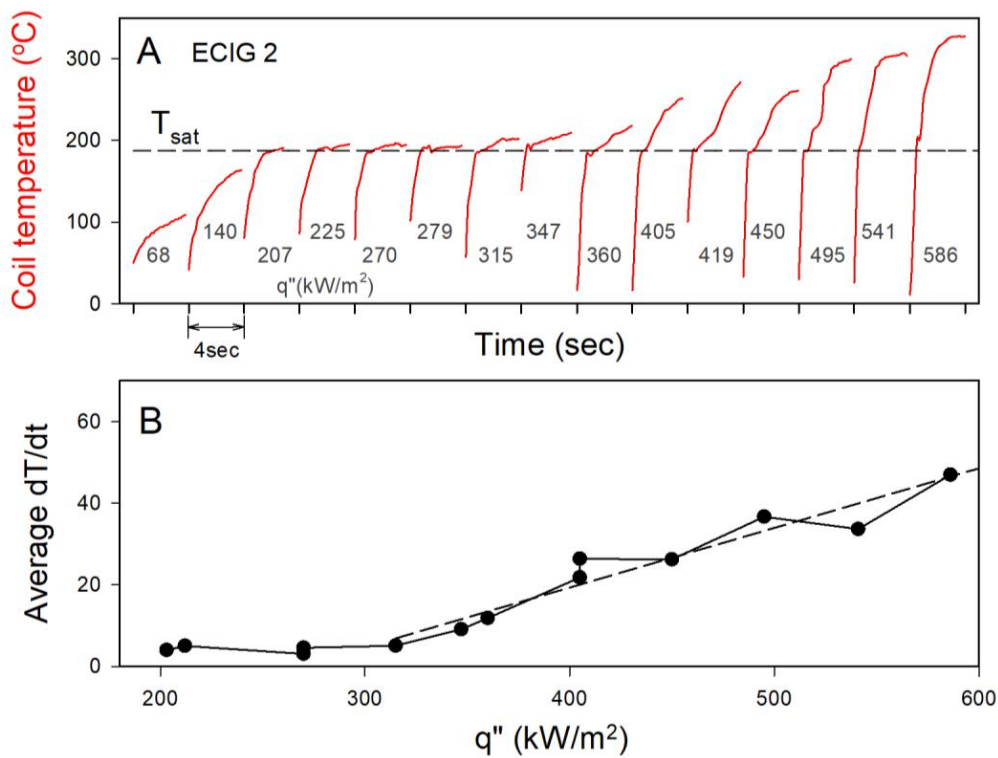

**Fig. 3 Temperature profile of a nickel-coil ECIG, during operation, at increasing heat fluxes.**
(A) Temperature profile for a sequence of repeated puffs, at stepwise increasing power for one SMOK TF-N2 ECIG. Note: the time axis is truncated between puffs for convenience of display. (B) Average rate of change in temperature for data in panel A, once the system reaches Tsat vs. q". dT/dt was calculated for all samples, excluding the transient region (rise in T prior to Tsat). Dashed line indicates best fit for linear portion of the rise in dT/dt.

Another window on film boiling in ECIGs is provided by observing the mass of liquid vaporized from the device as power input is increased. In particular, if a transition to film boiling occurs, the insulting effect of the vapor layer forming around the coil will impede further heat transfer to the liquid, and therefore result in diminishing returns in the amount of liquid vaporized with additional increments in q". The mass of liquid vaporized versus q" is shown in Figure 4 for the three ECIG devices examined in this study. For all three devices, at the lower heat fluxes, we observed a linear relationship between heat flux and mass of liquid vaporized; the greater the input, the proportionately greater the liquid vaporized. For heat fluxes greater than CHF, we found that additional heat flux resulted in diminishing returns in liquid vaporized. For example, for the SMOK TF-N2 device, a 50% increase in heat flux from 200 to 300 kW/m$^2$ resulted in an approximate 50% increase in liquid vaporized, while increasing q" from 400 to 600 kW/m$^2$ resulted in no significant increase in liquid vaporized. In this mode of operation, a greater fraction of the heat input goes into heating the wire itself and into greater conduction losses through the wire leads. Importantly, this plateau occurs within the manufacturer's recommended operating ranges for the VF Platinum and the SMOK V12-Q4, shown in shaded gray (there are no power guidelines for the SMOK TF-N2).

Also shown in Figure 4 are the CC yields per 15 puffs. Importantly, we found that the sharp increase in CC yields coincided with the plateau in liquid vaporized. Because carbonyls are thermal degradation products of PG and VG, (*12, 21, 28*) a sharp rise in carbonyls is consistent with a rise in the coil temperature; for the SMOK TF-N2 device, we indeed found that the superheat regime coincided with the rise in carbonyls. Table S4 includes a summary of the data.

In summary, we found that when CHF was exceeded, the ECIG coil temperature exceeded the PG boiling temperature, the ECIG exhibited a plateau in vaporization rate, and CC yields rose sharply. Importantly, the manufacturer's recommended windows of operation for the SMOK V12-Q4 and VaporFi devices exceed the CHF. Thus it is apparent that ECIG devices operating with PG liquid can exhibit film boiling behavior at plausible powers, and that onset of high carbonyl emissions is coincident with the onset of the transition to film boiling.

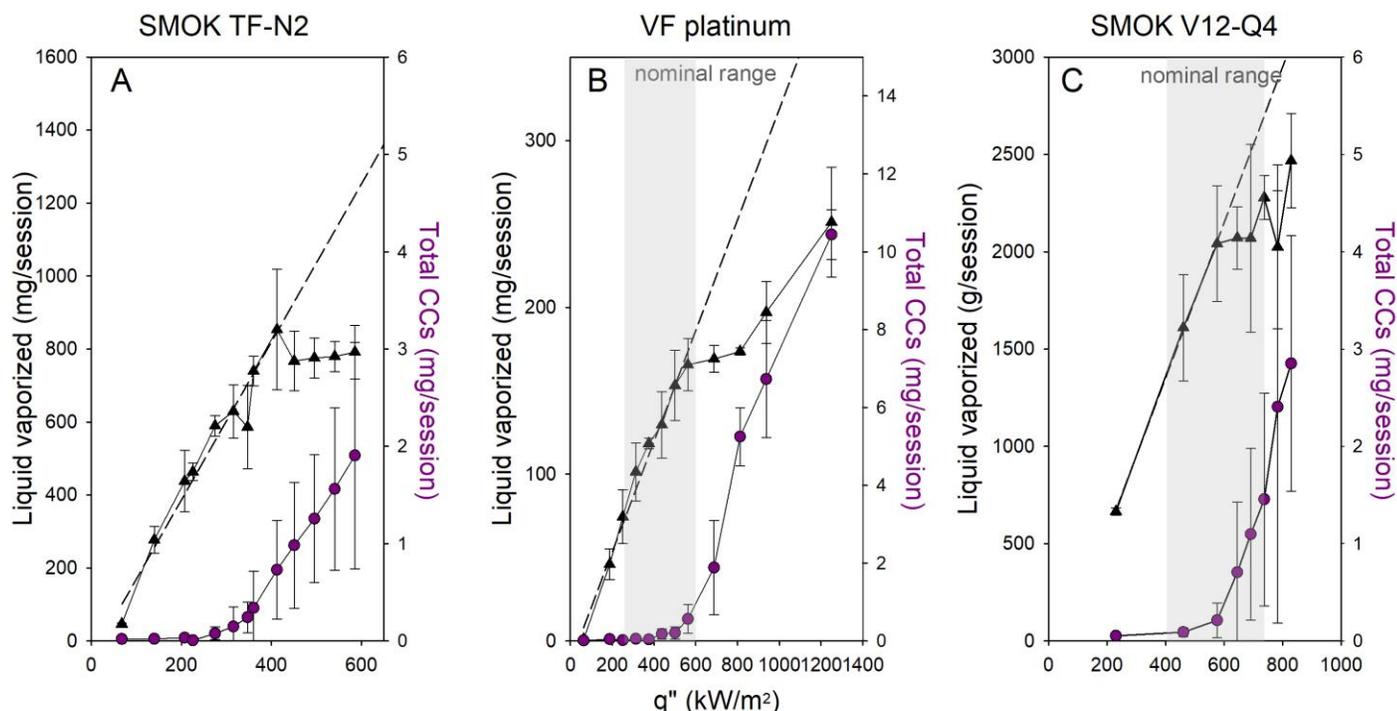

**Fig. 4 Liquid vaporized and total carbonyl yields vs. q" for the examined ECIG devices**
Mean (STD) liquid consumed and total carbonyl yields vs. q" (N=3 per condition). (A) SMOK TF-N2, (B) VaporFi platinum, and (C) SMOK V12-Q4. Dashed line indicates best fit for linear portion of the vaporization curve. Shaded area indicates the q" range corresponding to manufacturer's recommended battery settings and device geometry; SMOK TF-N2 has no manufacturer recommendations (Table S1).

## How is film boiling behavior affected by liquid composition?

Vegetable glycerin (VG) is the other major solvent used in ECIG liquids. Relative to PG, VG has a greater molar mass, and exhibits a higher boiling temperature, surface tension, heat of vaporization, and viscosity (Table S2). The Zuber equation (*29*) indicates that a heated surface submerged in VG should therefore exhibit a 50% greater CHF than if it were submerged in PG.

We repeated the submerged nickel wire measurements using VG liquid and found that the system transitioned to film boiling at much higher heat fluxes than PG, and did so in a manner that resulted in a catastrophic wire burnout; it was impossible to sustain a complete puff with a flux equal to or greater than CHF. We found the mean CHF for nickel wires submerged in VG to be equal to 866kW/m$^2$ (Fig. S5), more than double the mean attained with PG (343 kW/m$^2$).

We also examined mass of liquid vaporized and CC emissions for the SMOK TF-N2 filled with VG, up the maximum power attainable with the EScribe power supply for this setup. Table S5 includes a summary of the data. We found that the liquid vaporized continued to increase monotonically throughout the tested power range. Figure 4 shows the striking difference between the vaporization behavior of the PG and VG cases; PG vaporization plateaued at 350 kW/m2, while VG vaporization continued to rise throughout the test range. Consistent with this finding, the temperature profiles of the heating coil of the SMOK TF-N2 device did not depart from the previously described boiling regime, namely that during each puff, the temperature rapidly rose to boiling, and then remained at boiling for the remainder of the puff (Fig. S6). Most importantly, we found that across the range of heat fluxes examined, CC emissions did not rise drastically, unlike the case with PG (Fig. 5B).

In summary, all the signatures previously attributed to film boiling disappeared when PG was substituted with VG, under otherwise identical conditions, and carbonyl emissions remained proportional to q".

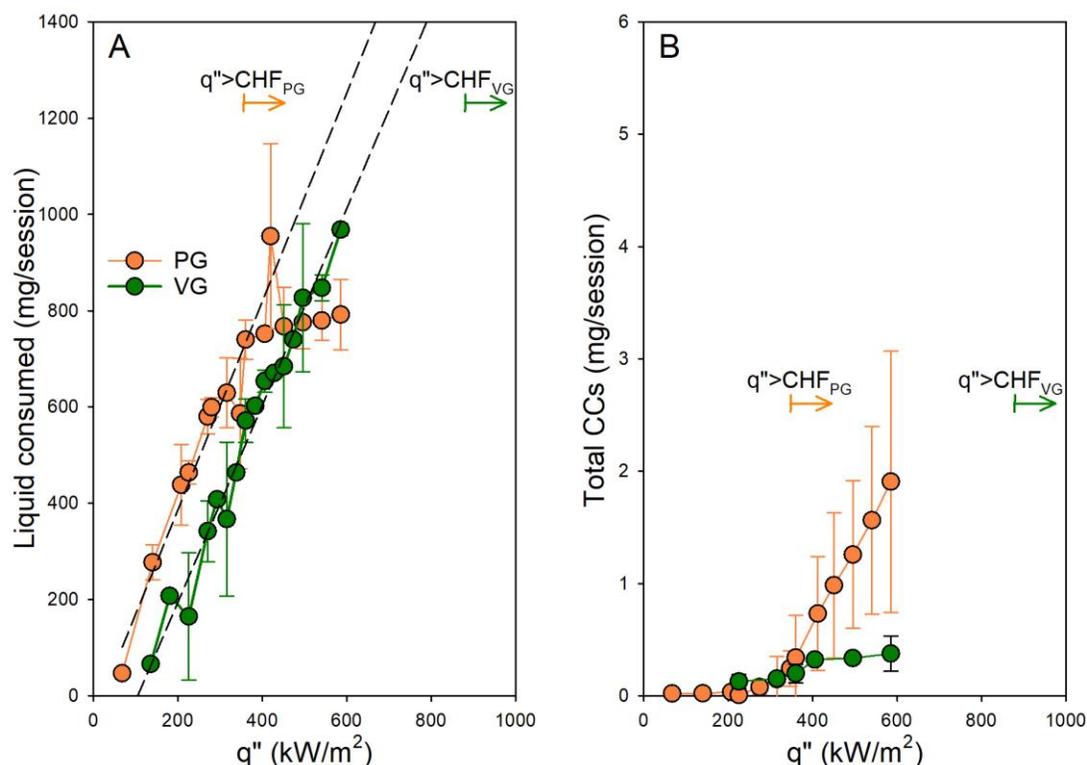

**Fig. 5 Effect of liquid composition on the film boiling behavior**
Mean (STD) liquid consumed (A) and total CCs (B) vs. q" (N=3 per condition) for the SMOK TF-N2 device filled with PG vs. VG. Dashed lines represents best fit linear regression for q"≤CHF

**Might limited vaporization, high temperatures, and high carbonyl emissions result from wicking limitations rather than film boiling?**

Many of the above reported observations could be explained by an alternative hypothesis that high carbonyl emissions result from dry puffs during which the liquid vaporization rate exceeds the rate at which fresh liquid can be wicked to the heating coil. For example, Figure 5 could indicate that the wick of the SMOK TF-N2 device transports PG at an intrinsically lower rate than VG, and that this rate-limiting process explains the observed vaporized liquid plateau for PG and not VG.

We compared the transport rates for PG and VG in vertically-suspended ECIG wicks. Figure 7 shows six IR images recorded at various points in time relative to the beginning of wick immersion in side-by-side beakers filled with PG or VG. The images show that PG always traveled a *greater* distance at a given time. Thus, the fact that VG vaporization continues to rise with q" while PG plateaus indicates that wicking transport limitations are not responsible for the phenomena observed when the CHF threshold is crossed.

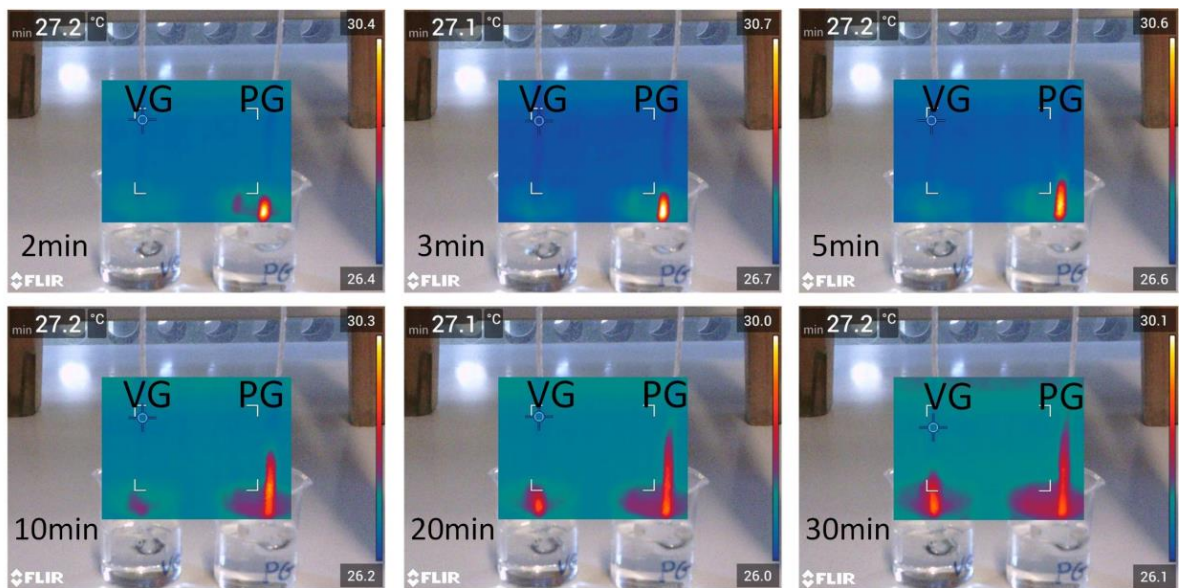

**Fig. 6 Comparison between the wicking rates of PG and VG.**
An IR camera continuously recorded the motion of PG and VG on silica wicks, partially submerged in PG and VG-filled beakers. The images show comparisons between the distances traveled by PG and VG with time.

**Discussion**

This study was conducted to understand whether the phenomenon of film boiling can explain the onset of a high carbonyl emissions regime during electronic cigarette operation. We examined submerged electronic cigarette heating coil wires and electronic cigarette devices for evidence of film boiling at realistic powers with PG and VG liquids and a variety of materials and geometries. We found that when ECIG devices are powered at heat fluxes that approached the CHF determined by direct observation of submerged heating wires, the devices exhibited temperature excursions above the boiling point, accompanied by a plateau in the liquid vaporization rate, and a concomitant rise in CC emissions. These observations could not be the result of the wick running dry during high power operation because when we replaced PG with the more viscous and more slowly wicking – but higher CHF – liquid, VG, all signs of film boiling disappeared, and carbonyl emissions dropped drastically. While manufacturers commonly instruct users to avoid high-temperature "dry puffs", during which the heating coil wick is insufficiently saturated, we found that manufacturer-recommend power ranges can lead to film boiling and excessive carbonyl emissions even when the wick is saturated. That is, when used as intended, ECIG devices can operate in a high-temperature film-boiling regime, leading to unnecessary exposure to toxicants. It has been argued that in real-world use, ECIG users detect and avoid high carbonyl emissions due to their associated unpleasant taste. (23) However, high levels of formaldehyde and other carbonyls can be found in the exhaled breath condensates of ECIG users, (*29*) indicating that ECIGs can expose users to high carbonyl emissions.

One limitation of this study is that only three devices were examined. To address this limitation, we compiled measurements of carbonyl emissions made by our group over the past 6 years for 25 commercially available devices (20 above-Ohm, and 5 sub-Ohm devices) that we characterized previously for power and heating coil surface area (range: 7-283 mm$^2$). Figure 7A shows total carbonyls vs. q" for 431 samples from 103 different conditions in which power ranged from 1 to 200 W, flow rates were set at either 1 or 1.5 L/min, and puff duration was 4 s. Consistent with the observations in this study, CC yields rose precipitously when q" approached a value of 450 kW/m$^2$. This value is near the mean CHF we found in the current study for nichrome heating wire with PG liquids. And as predicted by the CHF measurements with VG, carbonyl yields with VG liquids did not begin climbing until much higher values of q" (circa 700 kW/m2). Therefore, at least qualitatively, we see that historical data are consistent with the findings of this study: carbonyl emissions rise precipitously when q" exceeds CHF, and VG can reach substantially larger values of q" than PG prior to exhibiting HCR.

These findings therefore suggest that film boiling is a concern for a wide range of devices, and that ECIG heat flux should be limited to below the CHF to avoid unnecessary carbonyl exposure by users. As we have seen, however, CHF can vary widely even when the liquid and the heated surface are nominally identical, potentially due to variability in surface finish, among other factors. It may be more practical to regulate the maximum heat flux to a conservative value that is well below the theoretical CHF. For the data set presented in Figure 7A, we computed the probability that in 15 puffs, a given q" would result in carbonyl emissions that would exceed those of a combustible cigarette, approximately 2 mg (*29*). It can be seen in Figure 7B that for q" below 400 kW/m$^2$, the probability was approximately zero. This may be a reasonable upper bound for regulation. For reference, based on our previous measurements of heating coil dimensions and maximum power output, (*30, 31*) the JUUL ECIG has a q" of approximately 215 kW/m$^2$.

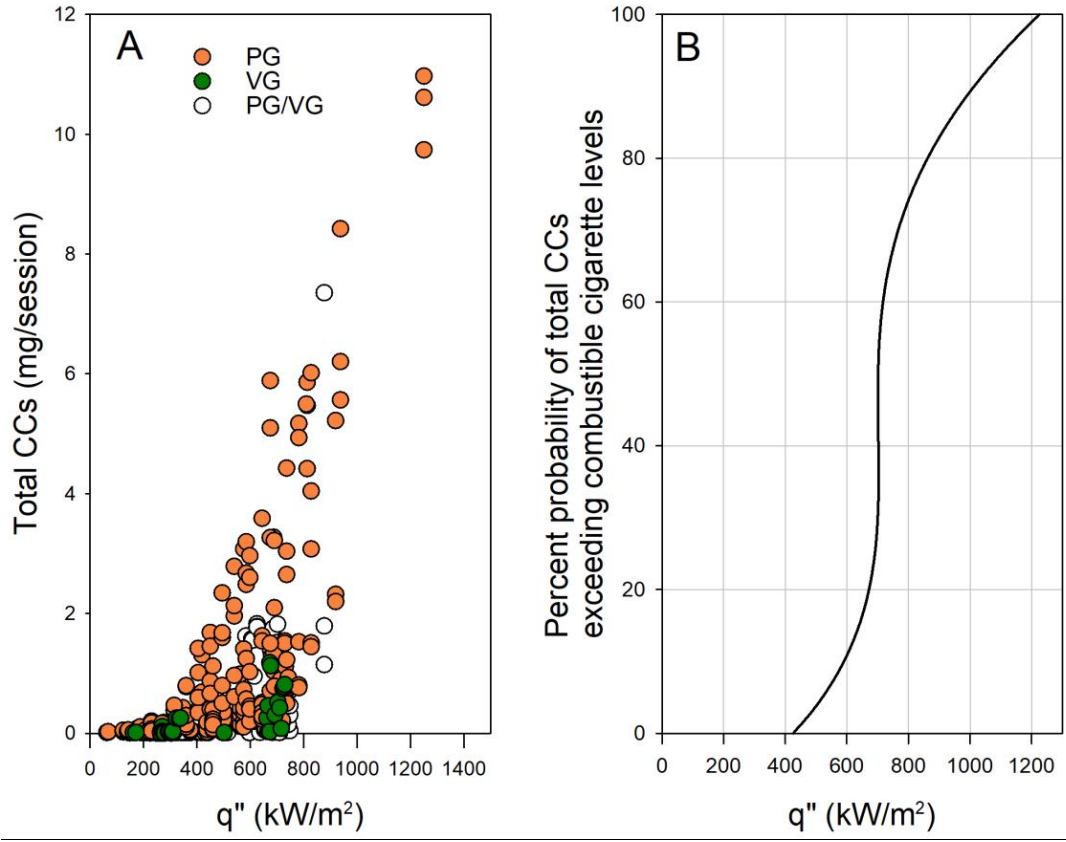

**Fig. 7 CC results from 431 aerosol samples that include 25 different ECIGs**
(A) Total carbonyls vs. q" for 25 ECIGs (N=431), 5 sub-Ohm and 20 above-Ohm devices, for powers ranging between 1-200W, ECIG coil surface area ranging between 7 and 283mm$^2$, and PG/VG:100/0-0/100. PG and VG are color-coded. (B) A model that shows the percent probability that ECIG total carbonyls exceed combustible cigarette levels for a given q". The model is built based on the empirical data in (A), smoothed using a 4$^{th}$-degree polynomial fit.

Another limitation of this study is that we examined only one class of toxicants produced by the pyrolysis of the ECIG liquids. The formation of other pyrolysis products found in ECIG aerosols, such as reactive oxygen species, (*4, 32-34*) VOCs,(*32*) and CO, (*35*) may become significant at heat fluxes lower than CHF. Nonetheless, this study has significant public health implications because it demonstrates that a simple quantitative design constraint on heat flux can greatly reduce emissions of a major class of respiratory toxicants. In summary, without compromising an ECIG's ability to deliver nicotine to smokers who want to switch, regulations that limit q" can reduce the toxicant load of ECIG aerosols.

**Acknowledgments:** None to declare**; Funding:** This work was supported by the National Institute on Drug Abuse (U54DA036105) and the Center for Tobacco Products of the US Food and Drug Administration; **Author contributions:** conceptualization: ST and AS, formal analysis: ST and AS, funding acquisition: AS, investigation: ST, RS, EK, MH, RH, AH, AS methodology: ST, RS, and AS, project administration: ST, RS, and AS, resources: RS, RH, NS, and AS, software: EK, and NK, supervision: ST and AS, validation: ST and AS, visualization: ST and AS, writing-original draft: ST, RS, EK, MH, RH, NK, AH, NS, and AS; **Competing interests:** AS is named on a patent application for a device that measures the puffing behavior of electronic cigarette users; and **Data and materials availability:** Key data supporting the findings in this study are provided in the main text and supplementary materials. Complete datasets may be requested from AS.

# Supplementary Materials for

## Hot wires and film boiling: Another look at carbonyl formation in electronic cigarettes

S. Talih, R. Salman, E. Karam, M. El-Hourani, R. El-Hage, N. Karaoghlanian, A. El-Hellani, N. Saliba, A. Shihadeh

Correspondence to: as20@aub.edu.lb

**This section file includes:**

    Supplementary text S1: Materials and Methods
    Supplementary text S2: EScribe Suite© validation
    Tables S1 to S5
    Figures S1 to S5
    Movie S1: Boiling regimes observed on a nickel wire submerged in a saturated PG solution
        Video link: https://youtu.be/3ZbJpmyKiAI

**Supplementary text S1: Materials and Methods**

This study was designed to test whether the onset of HCR in a given ECIG is predicted by exceedance of the CHF value that is intrinsic to the material and liquid combination of that ECIG and whether an ECIG operated at powers approaching and exceeding the predicted CHF threshold would exhibit vaporization behavior and temperature rise characteristic of transition to film boiling.

The study, therefore, involved measuring CHF for several ECIG heating coil wires submerged in a heated bath of PG and measuring EGIG aerosol emissions, namely liquid vaporized and carbonyls, as a function of heat flux. The resulting data were analyzed to test whether approaching or exceeding CHF resulted in both HCR onset and limited vaporization regime, in which additional increments in q" do not result in greater liquid vaporization rate due to the insulating effect of the vapor film around the coil. A plateau in vaporization rate with increasing q" is a hallmark of film boiling.

We repeated the above measurements using VG instead of PG liquid, and checked whether, as predicted by film boiling theory, VG exhibited higher CHF due to its higher density and surface tension. We also tested whether the heat flux at which HCR onset occurs and at which additional increments in q" result in diminishing amount of liquid vaporized shifted to proportionally higher values relative to the case for the PG liquid. Additionally, we compared the transport rates of PG and VG in vertically-suspended ECIG wicks to verify that the differences between PG and VG are due to differences in CHF and not differential wicking rates. The measurement procedures are described below.

Materials
Heating coil wires. We procured ECIG heating wires from online vendors in the USA. We selected kanthal, nickel, and nichrome wires of various thicknesses (22-32 gauge). We chose these specific metals because popular ECIG devices use kanthal and nichrome(*31, 36*), and ECIG aficionados that build their own coils also favor kanthal and nichrome (*37*). ECIGs that feature a temperature-control unit commonly use nickel due to its high-temperature coefficient of resistance (TCR), which is high enough to allow change in wire resistance to be employed as a temperature sensing principle. This property also allowed us to monitor the mean wire temperature directly.

Electronic cigarettes. We procured three ECIGs from internet vendors in the USA: SMOK®TF-N2 Air Core, VaporFi Platinum™ (VF), and SMOK® V12-Q4 (Fig. S1). These devices include both the sub-Ohm and above-Ohm categories of ECIGs. The SMOK®TF-N2 Air Core (0.15Ω) is a sub-Ohm device that features two nickel coils compartmentalized in two air passage holes. The VF platinum device (above-Ohm, 2.2Ω) is a standard bottom-coil tank that uses a nichrome coil, while the SMOK V12-Q4 (sub-Ohm, 0.15Ω) features four vertical kanthal coils connected in parallel and placed in a single air hole compartment. We disassembled samples from all ECIGs and measured the total coil surface area (Table S1).

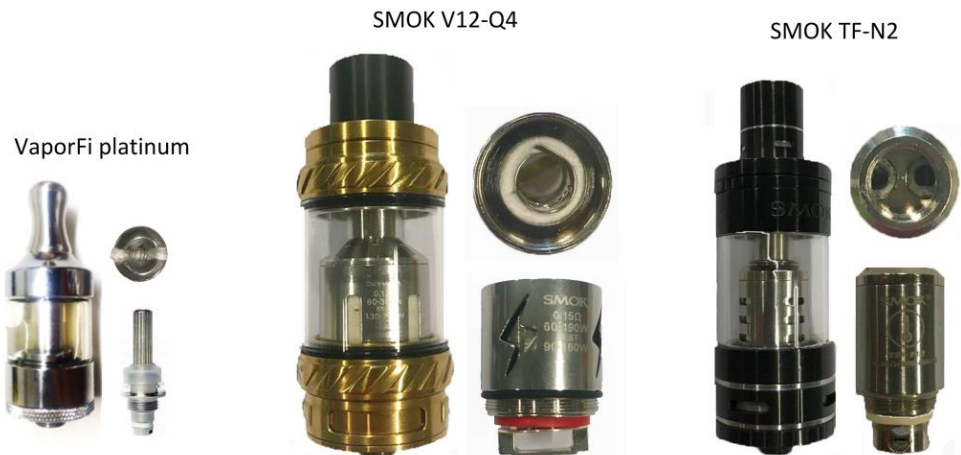

**Fig. S1** The ECIG devices used in this study

**Table S1.** ECIG features and dimensions

| ECIG | R (Ω) | Coil orientation | Wire metal | Wire diameter (mm) | Wire length (mm) | Number of wires | Total surface area (mm$^2$) | Recommended q" range (kW/m$^2$)[a] |
|---|---|---|---|---|---|---|---|---|
| SMOK TF-N2 | 0.12 | vertical | nickel | 0.27 | 131 | 2 | 222 | NA |
| SMOK V12-Q4 | 0.15 | vertical | kanthal | 0.36 | 52 | 4 | 235 | 415-737 |
| VF Platinum | 2.2 | horizontal | nichrome | 0.14 | 36 | 1 | 16 | 266-600 |

[a] the nominal q" ranges were calculated based on the manufacturer's recommended power range for the SMOK V12-Q4 (90-160W) normalized by the SMOK V12-Q4 coil surface area, and on the power calculated based the voltage range of a variable voltage battery sold by VaporFi (VaporFi Rocket: 3.2-4.8V) and the VaporFi coil surface area. NA: not applicable.

Liquids. Pure PG (99.5%) (CAS No 57-55-6) and VG (99–101%) (CAS No 56-81-5) were procured from Sigma-Aldrich Corporation.

Determination of CHF using submerged ECIG wires

To measure CHF in PG liquid, we placed a PG-filled glass beaker on a temperature-controlled hot plate (Corning™ PC-420D with 6795PR controller) and heated it to 188°C under atmospheric pressure (boiling temperature of PG; (*38*) while continuously monitoring the liquid temperature using a rapid response Type-K thermocouple probe (Omega Engineering, CT, USA) (Fig. 2A). We then individually submerged six nickel wires that differed by surface area (122-188m2) (Table S2) in the saturated PG solution and powered them using a 12V power supply controlled by a DNA200 circuit board (© 2018 Evolv LLC). We used adapters, removed from a rebuildable dripping atomizer device, to connect each nickel wire to copper leads, that are in turn connected to the DNA200 circuit board (Fig. 2A), used to collect data on the EScribe Suite (© 2018 Evolv LLC). The data collected include time, voltage, power, resistance, current, and temperature. In a side study, we verified the validity of the data reported by EScribe (Supplementary text S1).

Next, we powered the submerged wires for 10 seconds at each power level, starting with 1W, followed by 5W and increasing in 5W increments thereafter until well beyond the power at which the temperature spiked. The different boiling regimes were visually observable throughout the heating process, including the transition to film boiling. A video captured during a typical observation is available (Video S1).

In addition to the measurements recorded on EScribe, we calculated the wire temperature based on the measured wire resistance using the temperature coefficient of resistance for nickel after correction for the resistance of the connection wiring. We used the corrected heating wire resistance to calculate the heat flux supplied to the wire as:

$$q'' = \frac{I^2 R}{\pi D L}; \text{Equation 1}$$

Where $I$ is the current (A), $R$ is the heating wire resistance (Ω), $D$ is the diameter (m) of the wire, and $L$ is the length (m).

The coil temperature superheat ($\Delta T_{excess}$) was computed as:

$$\Delta T_{excess} = T_{coil} - T_{sat}; \text{Equation 2}$$

To measure the CHF for low TCR materials, we submerged fifteen kanthal (20-217mm$^2$) and 12 nichrome (31-228mm$^2$) wires (Table S3) in PG and powered the wires at increasing increments. Lab personnel observed the wires throughout the heating process and logged the power at which the wire began to glow. This critical power is denoted as $P^*$.

ECIG aerosol generation and sampling

The SMOK®TF-N2 Air Core ECIG device was filled with PG or VG and puffed using ALVIN, the AUB Aerosol Lab Vaping Instrument (*1, 38*), which was programmed to draw four-second puffs at a flow rate of 1LPM, with a 10-sec interpuff interval (*39*). An external power supply, coupled with a DNA200 circuit board, was used to supply power to the coil head. The power was increased in small increments over the range 15 to 130W. Data including voltage, power, current, and resistance were recorded using EScribe. We calculated the temperature using the temperature-resistance relationship of nickel. We accounted for the internal resistance of the ECIG connections by replacing the heating coils with a negligible resistance copper wire and measuring the ECIG resistance. We found it to be equal to 8mΩ.

The mouth end of the ECIG was connected to a filter holder fitted with a Gelman type A/E 47 mm glass fiber filter. Downstream of the filter holder was a 2,4-dinitrophenylhydrazine (Lp-DNPH) cartridge. The setup is described in detail in Talih, Balhas, Salman, Karaoghlanian and Shihadeh (*28*). Carbonyls were quantified by extracting the DNPH cartridges in 90/10 (vol/vol) ethanol (≥99.8%, CAS 64-17-5)/acetonitrile (≥99.93%, CAS 75-05-8) and analyzing the prepared extract by HPLC-UV, as described in Al Rashidi, Shihadeh and Saliba (*40*). The species analyzed, and the limits of detection and quantitation were, respectively (µg): formaldehyde, 0.002 and 0.007; acetaldehyde, 0.004 and 0.012; acetone, 0.001 and 0.004; acrolein, 0.003 and 0.012; propionaldehyde, 0.008 and 0.028; benzaldehyde 0.009 and 0.029; valeraldehyde, 0.002 and 0.007.

Measurements of total particulate matter (by pre- and post-weighing the glass fiber filter in its holder; Talih el al., 2017) and carbonyls were made in triplicate, each time using a new ECIG atomizer, as in Talih, Salman, Karaoghlanian, El-Hellani, Saliba, Eissenberg and Shihadeh (*16*). Before collecting aerosol, we performed a three-step cleaning and conditioning protocol. First, we programmed the puffing machine to draw clean air through the ECIG for 300 seconds without power supplied to the ECIG. Second, we programmed the puffing machine to generate 15 4-second puffs, again without powering the ECIG, and collected filter blanks and blank DNPH cartridge samples. Finally, we pre-conditioned each ECIG by drawing 15 puffs at 50W; the filter pads were replaced following the pre-conditioning step. Throughout all sessions, we maintained the liquid level in the ECIGs above the wicking holes and set the ECIG devices' airflow slot at half-open.

To avoid overloading the fiberglass filters, the number of puffs drawn differed according to the power input and ranged between 2-15puffs per session; we report the results on an equivalent 15-puff basis by scaling linearly. Table S4 and S5 provide a detailed list of conditions. Except for temperature measurement, the same procedures were repeated for VaporFi Platinum™ (VF) and SMOK® V12-Q4 (Fig. 4B) electronic cigarettes. The power was increased in small increments for each ECIG and ranged between 1-15W for the VaporFi platinum and 50-180W for the SMOK V12-Q4 (the wattage ranges recommended by each manufacturer). Each ECIG was pre-conditioned by drawing 15 puffs at 3W for the VaporFi platinum, and 50W for SMOK V12-Q4 device.

Theoretical CHF

We used the Zuber correlation (*41*) to calculate the theoretical CHF:

$$CHF = C\rho_g h_{fg} \left[\frac{g\sigma(\rho_f - \rho_g)}{\rho_g^2}\right]^{1/4}, \text{Equation 3}$$

Where the constant $C$ is 0.131, $\rho_g$ is the gas density (kg/m³), $h_{fg}$ is the latent heat of vaporization (J/kg), $g$ is the gravitational acceleration (m/s²), $\sigma$ is the surface tension (N/m), and $\rho_f$ is the fluid density (kg/m³) (Table S2).

**Table S2. Thermodynamic properties of PG and VG**

| Properties | PG | VG |
|---|---|---|
| Boiling point (°C) | 188 (*38*) | 293 (*38*) |
| Molar mass (g/mol) | 76.09 (*42*) | 92.09 (*43*) |
| Liquid density (kg/m³) | 870 (*44*) | 1261 (*44*) |
| Gas density (kg/m³) | 2.04 | 2.02 |
| Surface tension (mN/m) | 23.1 (*45*) | 51.9 (at 150°C) (*46*) |
| Latent heat of vaporization (kJ/kg) | 914 (*47*) | 974 (*47*) |

All properties are at saturation unless stated otherwise. The gas density is calculated based on the ideal gas equation.

Comparison between PG and VG transport rates in ECIG wicks

To measure the difference in transport rate between PG and VG, we submerged the end of silica wicks (3mm diameter, Lightning Vapes) in individual PG and VG-filled beakers and continuously monitored the rate at which the liquid traveled the wick using an infrared (IR) camera (FLIR Exx-Series, FLIR Systems) over a duration of 30 minutes. Weights were suspended from the submerged wick ends to insure that the wicks were vertically aligned. The difference in emissivity between liquid and wick allowed the position of the liquid front to be determined at each time interval.

**Supplementary text S2**: **EScribe Suite© validation**

EScribe Suite (Evolv LLC) is a software application for configuring, recording, and modifying the operation of DNA™ circuit boards. In this study, we collected data from EScribe and validated them using external measurements. Data include voltage, resistance, and temperature.

a. **Voltage**

We connected a kanthal coil of fixed resistance (0.4Ω) to a DNA250 chip, powered via a DC power supply (Fig. S2 left), and systematically increased the power on EScribe (1-30W in 5W increments). During operation, we continuously measured the voltage difference across the terminals of the DNA250 via a data acquisition system (DAQ, NI USB-6001). We found the voltage results from the two methods to be highly correlated ($R^2$=0.99) (Fig. S2 right), indicating that the EScribe voltage data are reliable.

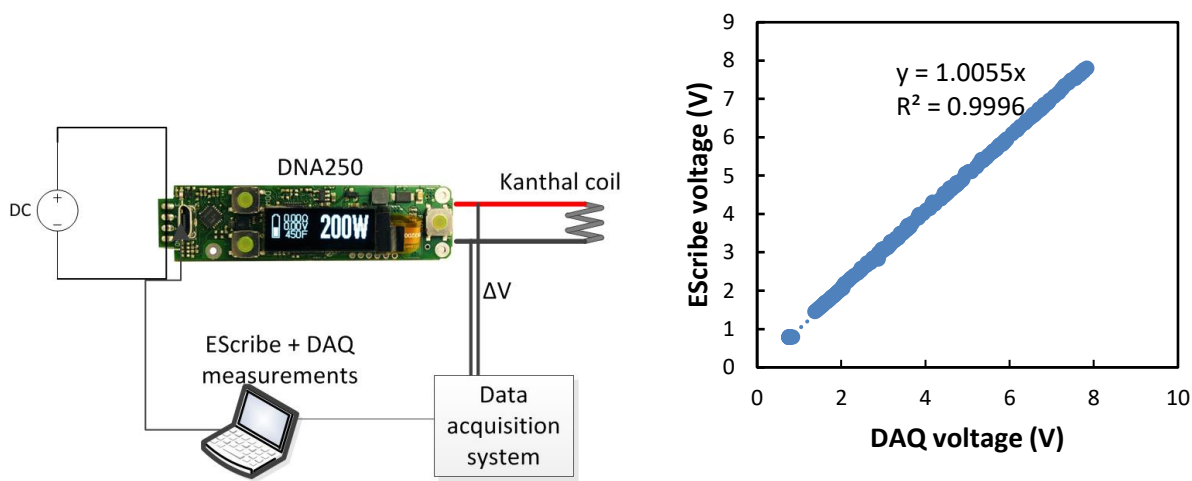

**Fig. S2** Left, Schematic of the setup used to verify the voltage set and measured by EScribe. Right, EScribe vs. DAQ voltage measurements

b. **Resistance**

We used a 50W-3Ω potentiometer (AVT05006E) and measured its resistance with a milliohm meter (GwInstek GOM-80). We then connected the potentiometer to a rebuildable dripping atomizer (RDA), fitted on a battery case with a DNA200 board (Fig. S3 left) coupled to a power supply, and measured its resistance on EScribe, via the 'Atomizer Analyzer' tool. We repeated the same procedure ten times, changing the position of the potentiometer wiper at every turn.

Figure S3 right shows a comparison between both measurements. We found an offset of approximately 0.2Ω, due to the added resistance of the RDA and electrical wires. However, after accounting for the added resistance, the EScribe data still overestimated the resistance measurements by 15%. These results indicate that resistance measurements from EScribe can be erroneous:
1) EScribe measures the overall resistance of the system, which often includes added resistance from the ECIG coil head and internal wiring.
2) Even when we subtracted the system resistance from the overall resistance, EScribe still overestimated the resistance by 15%.

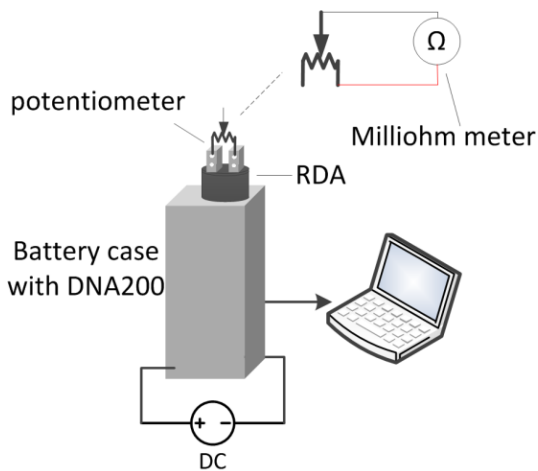
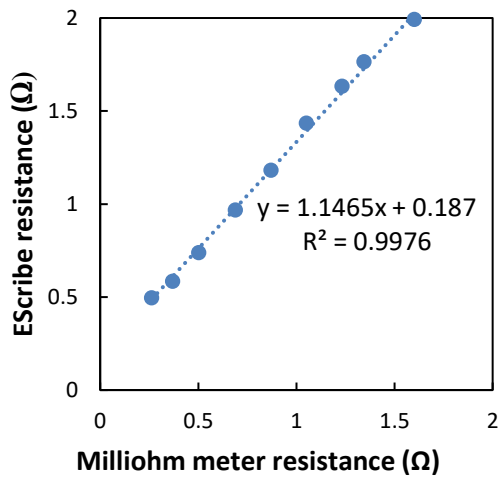

**Fig. S3** Left, schematic of the setup used to verify the resistance measured by EScribe. Right, EScribe vs. milliohm meter resistance measurements

### c. Temperature

We placed a water-filled bowl on top of a heated plate (Corning PC-420D) and submerged a nickel wire, fitted on the RDA/battery (similar to the setup described above) in the water. To reduce resistance from added connections, we submerged the RDA/wire system directly in the water-filled bowl, as in Fig. S4.

We then powered the wire at 1W for 4-5 minutes and slowly increased the temperature of the water (26-69°C) via the heater. Powering the coil at 1W was necessary to activate the coil and EScribe measurements, while also not significantly affecting the coil temperature.

During operation, we continuously measured:
a) The temperature of the liquid via a K-type thermocouple, connected to a DAQ (cDAQ1mod, NI9211, 1.25Hz).
b) The coil resistance on EScribe ("Live Ohm" readings), and converted the resistance to temperature based on the nickel temperature coefficient of resistance.

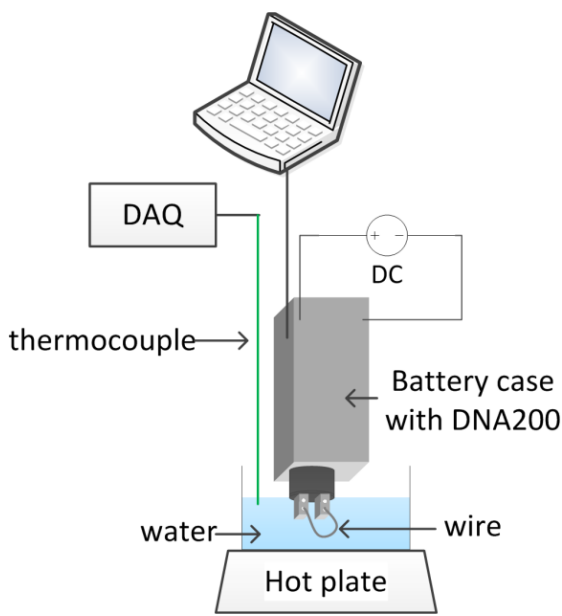

**Fig. S4** Escribe temperature validation setup

We found that after correcting for the internal resistance of the connection wiring, the thermocouple recorded and Escribe recorded data were in agreement to within 90% ($R^2 = 0.96$, $p<0.001$). These results indicate that if the system resistance is taken into account and subtracted from the resistance measurements collected on EScribe, EScribe temperature measurements are reliable.

**Table S3.**

Submerged wire conditions (wire metal, dimensions, and resistance) and results (P* and CHF)

| Metal | Wire diameter (mm) | Wire length (mm) | Number of coils | Surface area (mm²) | R (Ω) | P* (W) | CHF (kW/m²) |
|---|---|---|---|---|---|---|---|
| kanthal | 0.4 | 31 | 1 | 40 | 0.3 | 22 | 560 |
| kanthal | 0.4 | 47 | 1 | 60 | 0.5 | 27 | 454 |
| kanthal | 0.4 | 62 | 1 | 80 | 0.7 | 37 | 463 |
| kanthal | 0.4 | 78 | 1 | 100 | 0.9 | 43 | 427 |
| kanthal | 0.4 | 93 | 1 | 120 | 1.0 | 70 | 583 |
| kanthal | 0.2 | 32 | 1 | 20 | 1.4 | 9 | 453 |
| kanthal | 0.2 | 64 | 1 | 40 | 2.9 | 21 | 523 |
| kanthal | 0.4 | 47 | 1 | 60 | 0.5 | 23 | 389 |
| kanthal | 0.4 | 47 | 1 | 60 | 0.5 | 31 | 518 |
| kanthal | 0.4 | 44 | 4 | 217 | 0.5 | 100 | 460 |
| kanthal | 0.4 | 44 | 4 | 217 | 0.5 | 104 | 478 |
| kanthal | 0.4 | 40 | 1 | 50 | 0.3 | 29 | 585 |
| kanthal | 0.4 | 50 | 1 | 63 | 0.4 | 39 | 615 |
| kanthal | 0.4 | 75 | 1 | 94 | 0.5 | 62 | 662 |
| kanthal | 0.4 | 100 | 1 | 126 | 0.7 | 62 | 492 |
| nichrome | 0.5 | 37 | 1 | 60 | 0.2 | 17 | 288 |
| nichrome | 0.5 | 62 | 1 | 100 | 0.3 | 33 | 326 |
| nichrome | 0.5 | 87 | 1 | 140 | 0.5 | 53 | 380 |
| nichrome | 0.5 | 115 | 1 | 181 | 0.4 | 58 | 321 |
| nichrome | 0.5 | 145 | 1 | 228 | 0.5 | 85 | 371 |
| nichrome | 0.5 | 50 | 1 | 79 | 0.2 | 29 | 374 |
| nichrome | 0.5 | 40 | 1 | 63 | 0.1 | 19 | 305 |
| nichrome | 0.4 | 100 | 1 | 126 | 0.9 | 60 | 475 |
| nichrome | 0.4 | 50 | 1 | 63 | 0.4 | 30 | 473 |
| nichrome | 0.4 | 40 | 1 | 50 | 0.3 | 24 | 487 |
| nichrome | 0.4 | 25 | 1 | 31 | 0.2 | 15 | 471 |
| nichrome | 0.4 | 75 | 1 | 94 | 0.7 | 41 | 431 |
| nickel | 0.4 | 150 | 1 | 188 | 0.1 | 32 | 256 |
| nickel | 0.4 | 150 | 1 | 188 | 0.1 | 43 | 342 |
| nickel | 0.4 | 100 | 1 | 126 | 0.08 | 89 | 474 |
| nickel | 0.4 | 100 | 1 | 126 | 0.08 | 45 | 241 |
| nickel | 0.4 | 100 | 1 | 126 | 0.08 | 59 | 468 |
| nickel | 0.4 | 100 | 1 | 126 | 0.08 | 35 | 276 |
| nickel$_{VG}$ | 0.4 | 150 | 1 | 188 | 0.1 | 156 | 832 |
| nickel$_{VG}$ | 0.4 | 150 | 1 | 188 | 0.1 | 169 | 900 |

P* for kanthal and nichrome are determined from visual observations and temperature measurements for nickel wires. For all measurements, P* is corrected for resistance wire losses.

Apart from the last two rows which represent the conditions and results for nickel wires submerged in VG, all data are those for wires submerged in PG.

**Table S4.**

Aerosol generation conditions (ECIG device and description, puff conditions) and results (liquid vaporized and total carbonyls). All sessions were generated using a 4sec puff duration, 10sec inter-puff interval, and 1Lpm flow rate, and repeated in triplicates, each time using a new ECIG atomizer.

| ECIG | Power (W) | q"(kW/m$^2$) | R (Ω) | Number of sessions | Number of puffs | Mass of liquid vaporized (mg/15puffs) | Total carbonyls (µg/15puffs) |
|---|---|---|---|---|---|---|---|
| TFN2 | 15 | 68 | 0.22 | 3 | 15 | 46.7(2.36) | 19.64(2.74) |
| TFN2 | 31 | 140 | 0.22 | 3 | 15 | 277.17(36.44) | 21.2(5.01) |
| TFN2 | 46 | 207 | 0.22 | 3 | 15 | 437.9(83.93) | 33.37(15.78) |
| TFN2 | 50 | 225 | 0.22 | 3 | 2 | 463.5(24.06) | 6.77(8.28) |
| TFN2 | 60 | 270 | 0.22 | 3 | 2 | 579.75(36.16) | 54.56(91.31) |
| TFN2 | 62 | 279 | 0.22 | 3 | 3 | 599(20.3) | 101.16(21.94) |
| TFN2 | 70 | 315 | 0.22 | 3 | 2 | 629.25(73.03) | 147.17(202.7) |
| TFN2 | 77 | 347 | 0.22 | 3 | 2 | 586.25(114.72) | 243.68(157.86) |
| TFN2 | 80 | 360 | 0.22 | 3 | 2 | 740(40.7) | 340.43(376.36) |
| TFN2 | 90 | 405 | 0.22 | 3 | 2 | 752(8.69) | 712.94(605.74) |
| TFN2 | 93 | 419 | 0.22 | 3 | 2 | 954.5(192.6) | 750.97(523.2) |
| TFN2 | 100 | 450 | 0.22 | 3 | 2 | 767.5(81) | 982.78(646.22) |
| TFN2 | 110 | 495 | 0.22 | 3 | 2 | 775.75(54.99) | 1256.38(656.7) |
| TFN2 | 120 | 541 | 0.22 | 3 | 2 | 779.75(41.26) | 1561.76(833.37) |
| TFN2 | 130 | 586 | 0.22 | 3 | 2 | 792(73.5) | 1906.48(1163.69) |
| V12Q4 | 50 | 230 | 0.15 | 3 | 2 | 665.25(18.2) | 53.08(16.06) |
| V12Q4 | 100 | 461 | 0.15 | 3 | 2 | 1610.25(273.55) | 90.12(44.63) |
| V12Q4 | 125 | 576 | 0.15 | 3 | 2 | 2042(296.99) | 212.95(179.53) |
| V12Q4 | 140 | 645 | 0.15 | 3 | 2 | 2070.75(159.38) | 705.23(722.08) |
| V12Q4 | 150 | 691 | 0.15 | 3 | 2 | 2069.5(482.34) | 1097(881.13) |
| V12Q4 | 160 | 737 | 0.15 | 3 | 2 | 2279(112.12) | 1455.24(1093.39) |
| V12Q4 | 170 | 783 | 0.15 | 3 | 2 | 2025(421.17) | 2405.18(2222.22) |
| V12Q4 | 180 | 829 | 0.15 | 3 | 2 | 2467.75(242.95) | 2852.1(1313.33) |
| VF | 1 | 63 | 2.2 | 3 | 30 | 0.97(0.58) | 12.34(2136.48) |
| VF | 3 | 188 | 2.2 | 6 | 15 | 45.93(9.23) | 46.97(3.07) |
| VF | 4 | 250 | 2.2 | 3 | 15 | 74.43(16.1) | 20.94(43.49) |
| VF | 5 | 313 | 2.2 | 3 | 15 | 101.37(17.4) | 55.8(1.64) |
| VF | 6 | 375 | 2.2 | 3 | 15 | 118.33(3.35) | 40.22(39.49) |
| VF | 7 | 438 | 2.2 | 6 | 15 | 129.53(19.87) | 179.5(15.13) |
| VF | 8 | 500 | 2.2 | 3 | 10 | 153.25(21.02) | 207.47(132.05) |
| VF | 9 | 563 | 2.2 | 3 | 5 | 165.6(15.63) | 566.13(159.38) |
| VF | 11 | 688 | 2.2 | 3 | 5 | 169.1(8.06) | 1883.19(365.48) |
| VF | 13 | 813 | 2.2 | 3 | 5 | 173.6(2.27) | 5247.2(1209.78) |
| VF | 15 | 938 | 2.2 | 3 | 5 | 197.1(18.52) | 6727.72(747.06) |
| VF | 20 | 1250 | 2.2 | 3 | 5 | 251.1(33) | 10442.25(1501.97) |

**Table S5.**

Aerosol generation conditions and results (liquid vaporized and total carbonyls) for SMOK TF-N2 ECIG devices filled with VG. All sessions were generated using a 4sec puff duration, 10sec inter-puff interval, and 1Lpm flow rate, and repeated in triplicates, each time using a new ECIG atomizer.

| ECIG | Power (W) | q"(kW/m$^2$) | R (Ω) | Number of sessions | Number of puffs | Mass of liquid vaporized (mg/15puffs) | Total carbonyls (μg/15puffs) |
|---|---|---|---|---|---|---|---|
| TFN2 | 50 | 225 | 0.22 | 3 | 3 | 145.5(73.3) | 116.46(53.4) |
| TFN2 | 70 | 315 | 0.22 | 3 | 2 | 374(95.73) | 141.8(11.82) |
| TFN2 | 80 | 360 | 0.22 | 3 | 2 | 505.5(45.4) | 186.52(82.47) |
| TFN2 | 90 | 405 | 0.22 | 3 | 2 | 609(76.6) | 300.82(14.77) |
| TFN2 | 110 | 495 | 0.22 | 3 | 2 | 704(48.4) | 317.36(15.35) |
| TFN2 | 130 | 586 | 0.22 | 3 | 2 | 902.25(101.32) | 352.59(157.78) |

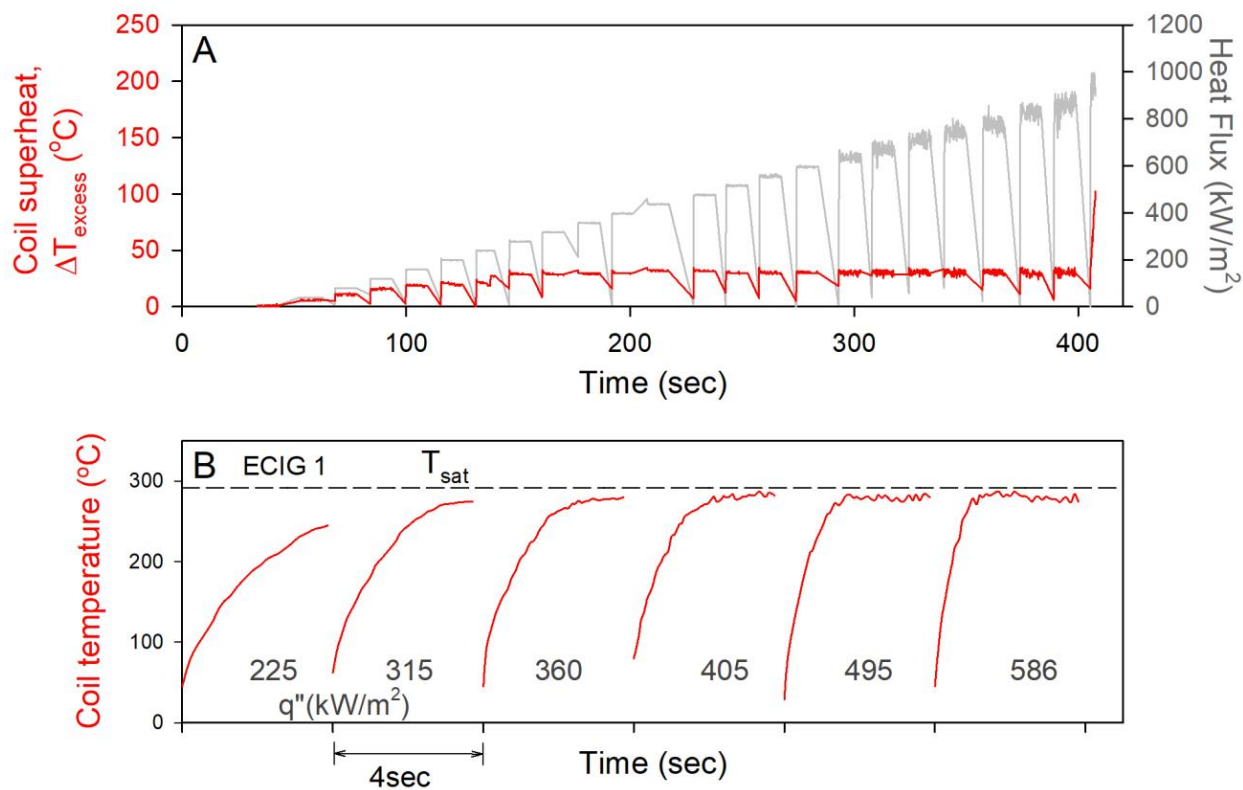

**Fig. S5 Critical heat flux determination for nickel wires submerged in VG**
(A) Example of the superheat profile vs. q" for one wire submerged in VG. (B) Temperature profile for a sequence of repeated puffs, at stepwise increasing power for one SMOK TF-N2 ECIG filled with VG liquid. Note: the time axis is truncated between puffs for convenience of display.